\documentclass{INTERSPEECH2023}

\usepackage{stfloats}
\usepackage{caption}
\usepackage{amsfonts}

\title{TranUSR: Phoneme-to-word Transcoder Based Unified Speech Representation Learning for Cross-lingual Speech Recognition}
\name{Hongfei Xue$^1$, Qijie Shao$^1$, Peikun Chen$^1$, Pengcheng Guo$^{1}$, Lei Xie$^{1*}$\thanks{*Corresponding author.}, Jie Liu$^2$}
\address{
  $^1$Audio, Speech and Language Processing Group (ASLP@NPU), \\
Northwestern Polytechnical University, Xi’an, China \\
  $^2$Huawei Cloud
  }
\email{hfxue@mail.nwpu.edu.cn, lxie@nwpu.edu.cn}
\usepackage{threeparttable}
\usepackage{multirow}
\usepackage{caption}
\begin{document}

\maketitle

\begin{abstract}
UniSpeech has achieved superior performance in cross-lingual automatic speech recognition (ASR) by explicitly aligning latent representations to phoneme units using multi-task self-supervised learning. While the learned representations transfer well from high-resource to low-resource languages, predicting words directly from these phonetic representations in downstream ASR is challenging. In this paper, we propose TranUSR, a two-stage model comprising a pre-trained UniData2vec and a phoneme-to-word Transcoder. Different from UniSpeech, UniData2vec replaces the quantized discrete representations with continuous and contextual representations from a teacher model for phonetically-aware pre-training. Then, Transcoder learns to translate phonemes to words with the aid of extra texts, enabling direct word generation. Experiments on Common Voice show that UniData2vec reduces PER by 5.3\% compared to UniSpeech, while Transcoder yields a 14.4\% WER reduction compared to grapheme fine-tuning.

\end{abstract}

\section{Introduction}
End-to-end ASR systems typically require a large amount of labeled data for training~\cite{20endtoendcomparison}, which is difficult to obtain in many languages. This problem makes training of ASR in most low-resource languages a challenging task~\cite{besacier2014automatic, 22Msurvey}. Cross-lingual learning is an approach widely used to improve ASR performance in low-resource languages~\cite{13cross, 18multilingual, 19largemulti}, which could learn common pronunciation information from non-target high-resource languages to assist ASR for target low-resource languages. Studies~\cite{conneau2019cross, 20unsupervised, 21XLSR, 21unispeech} have shown that pre-training on multilingual data, including labeled and unlabeled data from available language, and subsequent fine-tuning on the target low-resource language with labeled data, is an effective cross-lingual learning method.

Most recent research on pre-training for cross-lingual ASR can be classified into two categories. The first category only includes unsupervised loss in the pre-training stage, such as~\cite{conneau2019cross, 20unsupervised,  21XLSR, 21unsupervised, 22XLS-R, 22S3net}. These methods use much inexpensive, unlabeled multilingual data to learn cross-lingual features. In XLSR \cite{21XLSR}, a notable representation in the field is presented, which involves pre-training a model using self-supervised loss on 53 languages. The second category involves incorporating a supervised loss during unsupervised pre-training~\cite{21unispeech, 21jointcpcctc, 22JUST}. During the pre-training stage, cross-lingual representations is enhanced using labeled data from high-resource languages. The UniSpeech~\cite{21unispeech} method combines a supervised CTC loss~\cite{20ctc} and a self-supervised contrastive loss~\cite{20wav2vec}  to enhance the quality of the learned representations. Similarly, the JUST method~\cite{22JUST} integrates a supervised RNN-T loss with two unsupervised losses. For ASR tasks, studies such as UniSpeech~\cite{21unispeech} and JUST~\cite{22JUST} have demonstrated that the above-mentioned second category typically achieves higher performance compared to the first category such as XLSR~\cite{21XLSR}.

The choice of modeling units in supervised loss during the pre-training stage is typically between multilingual grapheme sets and phoneme sets. Some methods, like JUST~\cite{22JUST}, use grapheme units, while others, like UniSpeech~\cite{21unispeech}, opt phoneme units. To the best of our knowledge, the UniSpeech method is currently state-of-the-art (SOTA) on the cross-lingual Common Voice dataset~\cite{19commonvoice}. During the pre-training stage, the methods using grapheme units face challenges in learning shared cross-lingual representations due to a lack of shared graphemes among different languages. Conversely, using phoneme units could make models more easily learn shared phonetic representations. International Phonetic Alphabet (IPA)~\cite{99ipda} is a typical cross-lingual phoneme set that is suited for this situation. During the fine-tuning stage, a phoneme-based model cannot directly generate words, and requires additional fine-tuning with grapheme units. However, it is also challenging to map phonetic representations to grapheme units accurately.
In addition, the quantizer in UniSpeech is required to generate phoneme representations, which is difficult because the quantizer lacks sufficient trainable parameters.

This study proposes a \textbf{Tran}scoder based \textbf{U}nified \textbf{S}peech \textbf{R}epresentation (TranUSR) learning for cross-lingual ASR. Our TranUSR is a two-stage approach that includes a pre-trained UniData2vec and a phoneme-to-word (P2W) Transcoder. Firstly, we introduce UniData2vec in UniSpeech \cite{21unispeech}, which is a Data2vec \cite{22data2vec} model. UniData2vec replaces UniSpeech's quantized discrete representaions with continuous and contextual representations from a teacher model, enhancing the self-supervised targets to accurately capture phonetic information. Moreover, the teacher model has the same number of parameters as the student model, resulting in complete targets that are closely related to phonemes. This facilitates the learning of a unified speech representation in supervised learning and self-supervised learning. Secondly, we introduce Transcoder, which is based on the phoneme-to-grapheme model \cite{20p2g, 20jp2g, 21cascadernnt}. The Transcoder includes various components, such as a probabilistic vector input component. With these components, the Transcoder can convert encoder's hypotheses to target language's words with the help of information at different granulates. More importantly, the Transcoder can be trained with large-scale, low-cost text data, making it an attractive solution for low-resource settings. Experiments on the Common Voice dataset show that UniData2vec outperforms the SOTA method with an average improvement of relative ${5.3\%}$ phoneme error rate (PER). Furthermore, Transcoder achieves an average improvement of relative 14.4\% word error rate (WER) compared to grapheme fine-tuning.

\vspace{-5pt}
\section{Method}
\begin{figure*}[ht]
\centering
\includegraphics[width=0.9\textwidth]{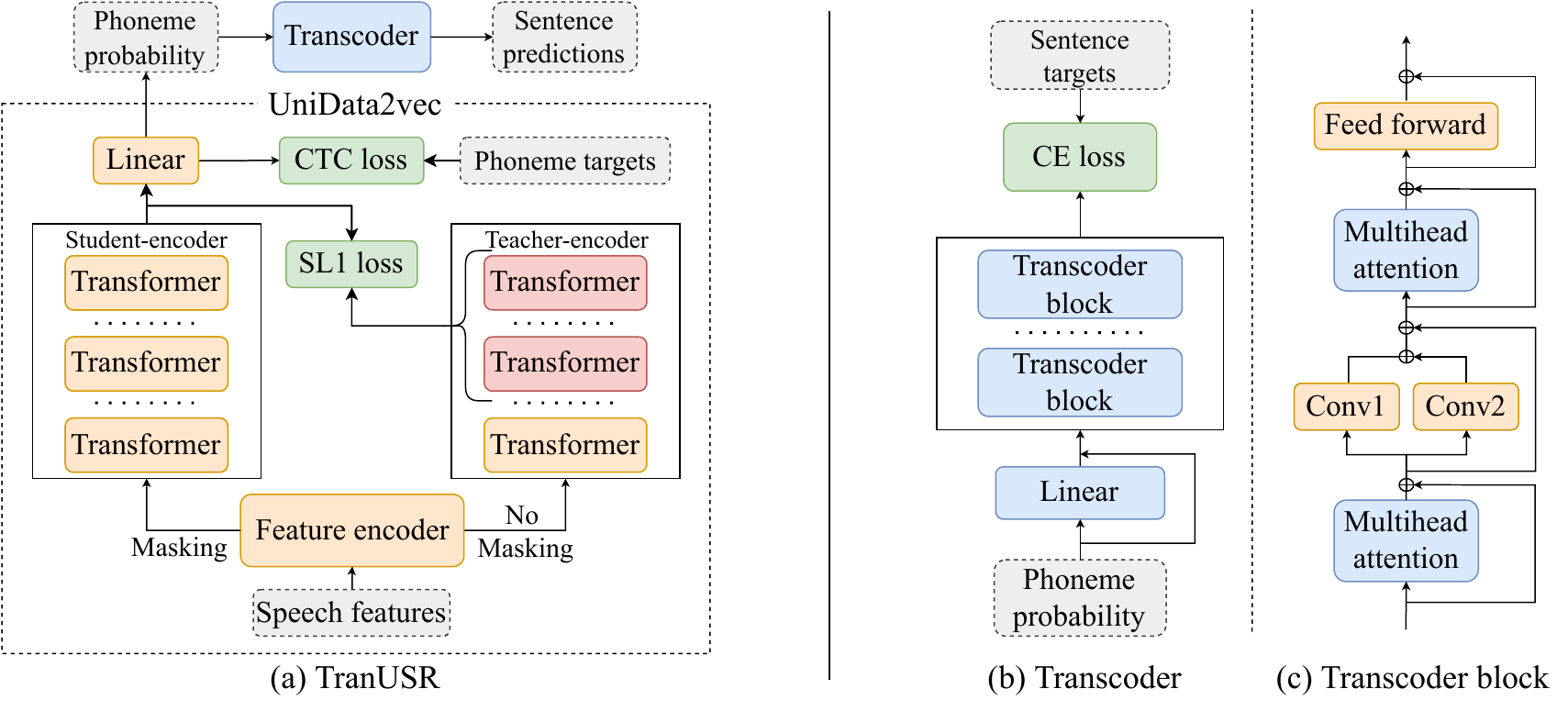}
\caption{\centering Architecture schematic diagram of TranUSR. Subfigure (a) is the overview of TranUSR, (b) is the Transcoder P2W module, and (c) is the details of the Transcoder blocks.}
\label{fig:TJSU}
\vspace{-5pt}
\end{figure*}

As illustrated in Figure~\ref{fig:TJSU} (a), the TranUSR framework comprises two modules: UniData2vec and Transcoder. UniData2vec predicts phoneme probabilities from speech features, and Transcoder for each language converts these probabilities into word-level sentences.
\vspace{-5pt}
\subsection{Problem statement}

In our research, we use four types of high and low-resource data, which can be denoted as $\mathbb{L}$, $\mathbb{U}$, $\mathbb{F}$ and $\mathbb{T}$. ${\mathbb{L}=\{(X^L_i, Y^L_i), i\in{(1, N_L)}\}}$ is labeled non-target high-resource language data consisting of speech audio and transcript labels. ${\mathbb{U}=\{(X^U_j), j\in{(1, N_U)}\}}$ is unlabeled target low-resource language data, ${\mathbb{F}=\{(X^F_k, Y^F_k), k\in{(1, N_F)}\}}$ is labeled target language data, and ${\mathbb{T}=\{(Y^T_l), l\in{(1, N_T)}\}}$ is pure text target language data. While $\mathbb{L}$, $\mathbb{U}$, and $\mathbb{T}$ are easy to obtain, $\mathbb{F}$ is expensive. 

Our task involves four steps: (1) training a UniData2vec base model on $\mathbb{L}$, (2) initializing with UniData2vec and training a UniData2vec$^+$ model on $\mathbb{U}$ to learn cross-lingual phonetically-aware representations, (3) fine-tuning the model on each target low-resource language dataset $\mathbb{F}$ to improve phoneme recognition ability, and (4) training a P2W Transcoder model for each language using the large text data $\mathbb{T}$.
\vspace{-5pt}
\subsection{UniData2vec}
\textbf{Model structure  } As shown in Figure.~\ref{fig:TJSU} (a), our UniData2vec module is composed of a convolutional feature encoder~\cite{20wav2vec}, two context Transformer encoders~\cite{17attention}, and a linear layer. The convolutional feature encoder 
is used to map the input speech $X$ to latent representations $Z$. The latent representations $Z$ are then fed into two Transformer encoders: a masked student encoder and an unmasked teacher encoder, which is similar to~\cite{22data2vec}.  The student encoder outputs context representations $C$, while the teacher encoder predicts targets $Y$. Finally, the context representations $C$ are mapped to phoneme units via the linear layer. In our method, the teacher and student encoders have the same amount of parameters, thus their phonetic representations have the same quality, which is more accurate than the quantizer outputs in UniSpeech~\cite{21unispeech}.

\textbf{Multi-task learning  } The multi-task learning of unified representation mainly consists of a supervised CTC loss~\cite{20ctc} and a self-supervised Smooth L1 (SL1) loss~\cite{22data2vec}. During this stage, our training object includes three parts: 1) Supervised phoneme recognition on the non-target language, which is achieved by using the CTC loss on ${\mathbb{L}}$ to align the context representations $C$ with phoneme units. 2) Self-supervised representation prediction on the non-target language, which is achieved by using the SL1 loss on ${\mathbb{L}}$ to encourage the student encoder to predict the masked samples'  targets $Y$. 3) Self-supervised representation prediction on the target languages, which is achieved by using the SL1 loss on ${\mathbb{U}}$.

\textbf{Self-supervised loss  } The self-supervised SL1 loss encourages the student encoder to predict the teacher's contextual representations $Y$ of masked samples. Specifically, the teacher encoder uses the exponential moving average (EMA)~\cite{22data2vec} method to parameterize the model parameters. The training targets $Y$ are constructed based on the output of the top $K$ layers of the teacher encoder, which represent the time steps masked in the student encoder. Given the context representations $C$ and the targets $Y$, we use the SL1 loss to regress these targets:
\begin{align}
L_{s}\left(Y, C\right)=\left\{
    \begin{array}{ll}
    \frac{1}{2}\left(Y-C\right)^{2} / \beta & \left|Y-C\right| \leq \beta \vspace{1pt} \\ 
    \left(\left|Y-C\right|-\frac{1}{2} \beta\right) & \text { otherwise }
    \end{array},
\right.
\end{align}
where $\beta$ controls the transition from square loss to L1 loss.

\textbf{Unified representation  } Supervised CTC and self-supervised SL1 losses are combined to train a unified representation. During supervised training, the primary loss is CTC, and the context representations $C$ generated by the student encoder correspond to phoneme units. Since the EMA method parameterizes the teacher encoder, the self-supervised targets $Y$ generated by the teacher encoder are also related to phoneme units. Therefore, the representations learned by supervised and self-supervised training can be constrained in the same space without the replacement in UniSpeech~\cite{21unispeech}. The losses can be expressed separately as: 
\begin{align}
  & L_{\text{UniData2vec}} = \sum_{(x,y)\in \mathbb{L}}(L_{\text{ctc}} + \alpha L_{\text{s}}) , \\
  & L_{\text{UniData2vec}^+} = L_{\text{UniData2vec}} + \sum_{x \in \mathbb{U}} L_\text{s} \ ,
\end{align}
where $\alpha$ is a tunable weight of self-supervised loss. In the training of dataset ${\mathbb{L}}$, it could be set to ${(0, 0.5)}$.

\textbf{Fine-tuning  } During the fine-tuning stage, we use the ${\mathbb{F}}$ data to train the CTC loss for phoneme recognition in each target language. This helps the model achieve better phoneme recognition ability on the specific language.

\subsection{Transcoder}
\textbf{Model structure  } As shown in Figure~\ref{fig:TJSU} (b), our Transcoder module consists of a linear layer and multiple Transcoder blocks. Each Transcoder block contains two convolutional layers, two multi-head attention blocks~\cite{17attention}, a feed-forward neural network, and multiple residual connections~\cite{16resnet}, as depicted in Figure~\ref{fig:TJSU} (c).

\textbf{Probabilistic vector input  } P2W models typically use phoneme one-hot vectors as input, which can result in performance degradation when phoneme sequences contain errors. Our Transcoder is trained on correct texts and fine-tuned on UniData2vec's hypotheses that may contain errors. To accommodate both types of input, we use soft phoneme posterior probabilities instead of hard one-hot vectors as input. When the phoneme sequence is from the correct text, its probability vector and one-hot vector are equivalent. However, the conventional text embedding layer only supports hard one-hot vectors as input. To overcome this limitation, we use a trainable linear layer to learn the text embedding, which enables correct text pre-training and hypothesis fine-tuning.

\textbf{Intra-word and inter-word information extraction  } Conventional P2W models typically use convolutional layers to extract intra-word phoneme dependencies. However, a single convolutional layer may not be sufficient to handle words of different lengths since they contain varying numbers of phonemes. To address this issue, our Transcoder module uses two convolutional models with different kernel sizes to extract intra-word information at two granularities, thereby improving accuracy for words of varying lengths. Additionally, we use a multi-head attention block to extract inter-word information.

\begin{figure}[ht]
  \centering
  \includegraphics[width=0.8\linewidth]{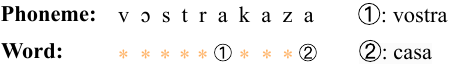}
  \caption{\centering{Sequence alignment using `$*$' tags as prefixes.}}
  \label{fig:Alignment Loss}
\end{figure}

\textbf{Alignment constraint loss  } The sequence lengths of phonemes, words can differ between languages, and CTC loss is often used to handle inconsistent output and input sequence lengths. However, P2W mapping has clear alignment boundaries, unlike audio-to-word mapping. To further improve the performance of Transcoder, we use cross-entropy (CE) loss to ensure accurate alignment learning. Since CE loss requires inputs and outputs to have the same length, we align the sequences by prefixing words with `$*$' tags. As an example, for the Italian sentence ``vostra casa" we align it as shown in Figure~\ref{fig:Alignment Loss}. With this alignment, our Transcoder can learn to accurately map phonemes to words.

\section{Experimental setup}
\textbf{Dataset  } This study uses two datasets: Common Voice 6.0 \footnote{https://commonvoice.mozilla.org/en/datasets. Dec 2020 release version is used for training our models. We use the same data size for each language compared to the UniSpeech.}~\cite{19commonvoice} and TED2020 \footnote{https://opus.nlpl.eu/TED2020.php}~\cite{20ted}. To compare with UniSpeech, we use the following eight languages for experiments: English (\textit{en}), Spanish (\textit{es}), French (\textit{fr}), Italian (\textit{it}), Kyrgyz (\textit{ky}), Dutch (\textit{nl}), Russian (\textit{ru}) and Tatar (\textit{tt}). We use en as a non-target language $\mathbb{L}$ with 1350 hours of data.  Others are the target languages, and the duration of $\mathbb{U}$ is in Table~\ref{tab:unidata2vec}. For fine-tuning, we use the same settings of  UniSpeech~\cite{21unispeech}. TED2020 is a multilingual text corpus, and we select the following three languages as $\mathbb{T}$ for evaluation with 400k sentences each: \textit{es}, \textit{fr} and \textit{it}.



\textbf{UniData2vec structure  } We implement the model within the fairseq~\cite{19fairseq} framework and refer to the settings in Data2vec2~\cite{22data2vec}. The encoder consists of 7 convolution layers, each with 512 channels, the strides are (5, 2, 2, 2, 2, 2, 2), the kernel widths are (10, 3, 3, 3, 3, 2, 2). Both the teacher and student encoders have 12 Transformer layers, whose dimensions are 768, inner dimensions are 3072, and have 8 attention heads. The targets are determined by taking the average of the top $K = 8$ layers. We generate IPA phoneme transcriptions by an open-source phonemizer tool \footnote{https://github.com/bootphon/phonemizer}.

\textbf{UniData2vec details  } During the pre-training stage, we set the weight $\alpha$ of the self-supervised loss to 0.15 and the weight $\beta$ in SL1 to 0.25. We first train a base UniData2vec model on ${\mathbb{L}}$ and then transfer it to ${\mathbb{U}}$ to form UniData2vec$^+$. It is worth noting that both our UniData2vec and UniData2vec$^+$ models have the same structure. In the fine-tuning stage, we followe the settings in~\cite{22data2vec}. We use Adam optimization~\cite{14adam} with a peak learning rate of $5\times 10^{-4}$ for UniData2vec and $5\times 10^{-5}$ for UniData2vec$^+$. The UniData2vec model is updated 400K times, while UniData2vec$^+$ is updated 200K times.

\textbf{Transcoder structure  } We use three Transcoder blocks with a model dimension of 256 and an inner dimension of 1024, four attention heads, and two convolutional kernels of sizes 3 and 5. For each language, we select the top 50,000 most frequent words as the output vocabulary.

\textbf{Transcoder details  } Transcoder is first trained on $\mathbb{T}$ and then optionally fine-tuned using UniData2vec's hypothesis. We use Adam optimization with a peak learning rate of 5e-4 for 200k updates and a batch size of 256 sentences.

\section{Results}
\subsection{UniData2vec}
\textbf{Cross-lingual phoneme recognition  }
As shown in Table~\ref{tab:unidata2vec}, Our re-implementation of the UniSpeech and UniSpeech$^+$ models shows some improvements and declines compared to the published UniSpeech~\cite{21unispeech} results, which could be attributed to the updates of the Common Voice dataset and the phonemizer tool. 
After being fine-tuned on low-resource language $\mathbb{F}$ for 1 hour, the UniData2vec shows similar competitiveness compared to the re-implemented UniSpeech. But when the fine-tuning dataset is added to 10 hours, the average PER of UniData2vec10 is reduced to relative ${5.8\%}$ compared with UniSpeech10. We think this is because UniData2vec gives more importance to supervised CTC loss as the primary loss, resulting in better phoneme recognition abilities, which require more data to be transferred to other languages.
When further training with the respective monolingual unlabelled data $\mathbb{U}_{\text{\textit{mo}}}$, UniData2vec$^+$ is on average 5.3\% lower than UniSpeech$^+$ in terms of PER. This improvement is more prominent in languages with abundant unlabeled data such as \textit{es}, \textit{fr}, and \textit{it} but weaker in languages with little data such as \textit{ky}, \textit{ru}, and \textit{tt}. Because the weight of self-supervised loss is small during UniData2vec, and the ability to predict masked positions is not enough, thus UniData2vec$^+$'s advantages require more data to reflect.

\begin{table*}
\centering
\caption{PER (\%) results on Common Voice for different methods. The $\mathbb{L}$ and $\mathbb{U}$ datasets are described in detail in Section 3. $\mathbb{U}_{\text{\textit{mo}}}$:  Monolingual unlabelled data. Re-implement: we re-ran the baseline experiment from UniSpeech.}
\label{tab:unidata2vec}
\renewcommand{\arraystretch}{0.8}
\resizebox{0.95\textwidth}{0.13\textwidth}{
\begin{tabular}{lcccccccccc}
\toprule
Model      & pre-train & fine-tune & \textit{es}   & \textit{fr}   & \textit{it}   & \textit{ky}   & \textit{nl}   & \textit{ru}   & \textit{tt}  & avg  \\ \midrule
Unlabeled data ($\mathbb{U})$   &  &    & 168 h & 353 h & 90 h  & 17 h  & 29 h  & 55 h  & 17 h &      \\ \midrule
UniSpeech ~\cite{21unispeech}  & $\mathbb{L}$  & 1 h & 10.9 & 14.8 & 15.2 & 11.4 & 16.2 & 16.1 & 9.6 & 13.5 \\
UniSpeech$^+$~\cite{21unispeech}  & $\mathbb{L}, \mathbb{U}_{\text{\textit{mo}}}$ & 1 h  & 5.7  & 7.9  & 8.1  & \textbf{6.8}  & 9.3  & \textbf{8.6}  & \textbf{6.0} & 7.5  \\ \midrule
UniSpeech (re-implement)   & $\mathbb{L}$   & 1 h  & 10.4 & 12.7 & 13.4 & 10.2 & 14.8 & 15.7 & 8.8 & 12.3     \\
UniData2vec & $\mathbb{L}$   & 1 h  & 10.0 & 13.0 & 13.3 & 10.3 & 15.0 & 15.1 & 9.0 & 12.2     \\
UniSpeech10   & $\mathbb{L}$   & 10 h & 7.2  & 8.1  & 10.9 & 4.3  & 7.3  & 7.1  & 3.3 & 6.9     \\
UniData2vec10  & $\mathbb{L}$   & 10 h & 7.0  & 8.2  & 10.4 & 4.1  & 6.2  & 6.6  & 3.2 & 6.5     \\
UniSpeech$^+$ (re-implement)   & $\mathbb{L}, \mathbb{U}_{\text{\textit{mo}}}$   & 1 h  & 6.1  & 7.7  & 8.5 & 7.8  & 10.8 & 9.8  & 7.4  & 8.3      \\
UniData2vec$^+$ & $\mathbb{L}, \mathbb{U}_{\text{\textit{mo}}}$   & 1 h  & \textbf{5.5}  & \textbf{6.5}  & \textbf{7.1}  & \textbf{6.8}  & \textbf{8.0}  & 8.8  & 7.2  & \textbf{7.1}     \\ \bottomrule
\end{tabular}
}
\vspace{-10pt}
\end{table*}

\begin{figure}[ht]
  \centering
  \includegraphics[width=\linewidth]{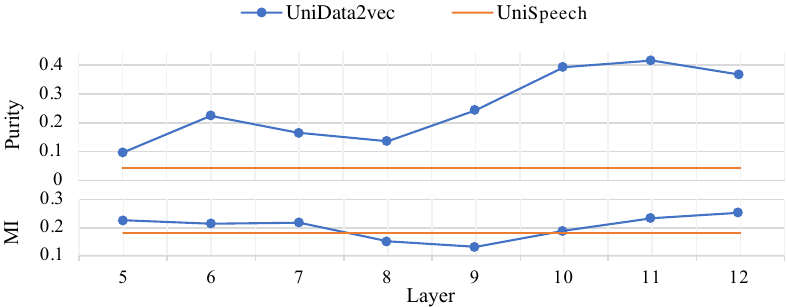}
  \caption{\centering Quality of cluster assignments measured by Purity and Mutual Information (MI) obtained through k-means clustering on self-supervised targets.}
  \label{fig:purity}
  \vspace{-10pt}
\end{figure}

 
\textbf{Unified representation learning  }
To demonstrate the superiority of UniData2vec's unified representation learning over UniSpeech, we evaluate the quality of the self-supervised phoneme information learned by UniData2vec. Following~\cite{21hubert}, we use $K$-means to cluster the self-supervised targets' phoneme information and measure purity and mutual information metrics. We extract representations from the top 8 layers in the teacher encoder for UniData2vec and use discrete latent representations for UniSpeech. Figure~\ref{fig:purity} shows that UniData2vec exhibits a clearer purity than UniSpeech, with the difference becoming more evident in deeper layers. This is because the final layer performs phoneme classification. While UniData2vec generally outperforms UniSpeech in terms of mutual information, there is a performance drop in layer 9, possibly due to the intermediate layers learning less content information.

\subsection{Transcoder}

\textbf{P2W transcoder vs. direct fine-tuning  } For a model like UniSpeech or UniData2vec that learns phoneme recognition in the pre-training stage, we can transfer it to predict words of target languages by directly replacing the output units and then fine-tuning. In order to demonstrate the advantages of the Transcoder, we compare it with direct fine-tuning. As shown in Table~\ref{tab:phoneme to word}, the Transcoder trained with extra text data results in significantly better results than the baseline BPE~\cite{15bpe} model, outperforming the grapheme model WER by about ${14.4\%}$. Pure text data is cheaper and easier to obtain than paired fine-tuning data, which makes the Transcoder scheme more practical.


\begin{table}[ht]
\centering
\caption{\centering WER (\%) comparison of Transcoder and the direct fine-tuning of UniData2vec$^+$. Both the two fine-tuning in the table using 1 hour of data. The BPE model has an output label of 4000 BPE units. The Grapheme model uses grapheme sets from the 400k sentences.}
\label{tab:phoneme to word}
\resizebox{1.0\linewidth}{!}{
\begin{tabular}{@{}lcccc@{}}
\toprule
Model & \textit{it} & \textit{es} & \textit{fr} & avg \\ \midrule
UniData2vec$^+$ + BPE fine-tune\tnote{\dag}            & 63.5  & 62.9     & 69.7      & 65.4 \\
UniData2vec$^+$ + Grapheme fine-tune\tnote{\ddag}        & 29.6  & 29.5     & 45.3      & 34.8 \\
UniData2vec$^+$ + Transcoder               & \textbf{27.4}  & \textbf{25.7}     & \textbf{36.2}      & \textbf{29.8}  \\ \bottomrule
\end{tabular}
}
\end{table}

\textbf{Ablation study of Transcoder  } We conduct ablation experiments to verify the role of each component in the Transcoder. As shown in Table~\ref{tab:ablation}, we evaluate the performance of our proposed Transcoder architecture by ablating its key components, including the use of probabilistic inputs, two convolution modules, and CE loss, using correct text and UniData2vec$^+$ hypothesis inputs as evaluation metrics. The results demonstrate that removing each of these components leads to weighted average WER increase of 7.3\%, 7.0\%, and 63.2\%, respectively. The Conformer architecture using CE loss compared to our Transcoder resulted in a higher weighted average WER of 6.3\%, demonstrating the effectiveness of our proposed architecture.

\begin{table}[ht]
\caption{Ablation study of each component's impact on WER (\%) of the Transcoder. Only Italian text is used in experiment.}
\label{tab:ablation}
\resizebox{1.0\linewidth}{!}{
\begin{tabular}{lccccc}
\toprule
\multirow{2}{*}{Input data} & \multicolumn{4}{c}{Transcoder}                   
& \multirow{2}{*}{Conformer} \\ \cline{2-5}
& Standard 
& \multicolumn{1}{c}{\begin{tabular}[c]{@{}c@{}}w/o \\ Prob inputs\end{tabular}} 
& \multicolumn{1}{c}{\begin{tabular}[c]{@{}c@{}}w/o \\ Two conv\end{tabular}} 
& \multicolumn{1}{c}{\begin{tabular}[c]{@{}c@{}}w/o\\ CE loss\end{tabular}} 
&   \\ \midrule
Correct text                & \textbf{3.2}     & 3.2      & 3.6     & 6.7  & 3.5   \\
Hypothesis                   & \textbf{27.4}     & 29.4   & 27.8  & 32.1 & 28.3   \\ \bottomrule
\end{tabular}
}
\vspace{-12pt}
\end{table}

\section{Conclusions}

This paper presents an effective approach for low-resource languages that utilizes unified representation learning on phonetically-aware representations, followed by a P2W Transcoder. The proposed UniData2vec model achieves an improvement of over 5.3\% compared to the SOTA method, while the Transcoder model yields a 14.4\% relative reduction in word error rate compared to grapheme fine-tuning. However, UniData2vec$^+$ showes weaker performance for low amounts of unlabeled data. In the future, we will unify unlabeled data from multilingual languages to pre-train.




\newpage

\bibliographystyle{IEEEtran}
\bibliography{TranUS}

\end{document}